\documentclass[twocolumn,showpacs,amsmath,amssymb,aps,superscriptaddress,pra]{revtex4}
\begin{document}

\title{Continuous variable teleportation as a generalized thermalizing
        quantum channel}

\author{Masashi Ban}
\affiliation{Advanced Research Laboratory, Hitachi, Ltd.,
             1-280 Higashi-Koigakubo, Kokubunji, Tokyo 185-8601, Japan}
\author{Masahide Sasaki}
\affiliation{Communication Research Laboratory,
             4-2-1 Nukui-Kitamachi, Koganei, Tokyo 184-8795, Japan}
\affiliation{CREST, Japan Science and Technology}
\author{Masahiro Takeoka}
\affiliation{Communication Research Laboratory,
             4-2-1 Nukui-Kitamachi, Koganei, Tokyo 184-8795, Japan}
\affiliation{CREST, Japan Science and Technology}

\date{\today}

\begin{abstract}
A quantum channel is derived for continuous variable teleportation
which is performed by means of an arbitrary entangled state
and the standard protocol.
When a Gaussian entangled state such as a two-mode squeezed-vacuum
state is used, the continuous variable teleportation is equivalent
to the thermalizing quantum channel.
Continuous variable dense coding is also considered.
Both the continuous variable teleportation and the continuous variable
dense coding are characterized by the same function determined
by the entangled state and the quantum measurement.
\end{abstract}

\maketitle

Quantum information processing attracts much attention
in quantum physics and information science
\cite{nielsen2000a,ekert2000a,alber2001a},
which gives deeper insight into the principles of quantum
mechanics and provides novel communication systems
such as quantum teleportation and quantum dense coding.
Quantum teleportation transmits an unknown quantum state
by means of classical communication and quantum entanglement
\cite{bennett1993}.
Quantum dense coding sends classical information
via a quantum channel and quantum entanglement
\cite{bennett1992}.
Bowen and Bose have recently shown that a finite dimensional
quantum teleportation with the standard protocol is equivalent
to a generalized depolarizing quantum channel \cite{bowen2001b}.
In this letter, we generalize this result to
infinite dimensional quantum teleportation and we show that
continuous variable quantum teleportation with the standard
protocol is equivalent to a generalized thermalizing channel.

We explain the standard protocol of continuous variable quantum
teleportation \cite{braunstein1998,furusawa1998}.
Alice and Bob first share an entangled quantum system in a quantum
state $\hat{W}^{AB}$ which is arbitrary though a two-mode
squeezed-vacuum state is usually used,
where $A$ and $B$ stand for the quantum system at Alice's and Bob's sides.
When Alice is given a quantum state $\hat{\rho}^{Q}$
to be teleported to Bob, she performs the simultaneous measurement
of position and momentum of the compound quantum system $Q+A$.
Then Alice sends the measurement outcome $(x,p)$ to Bob
via a classical communication channel.
After receiving the measurement outcome,
Bob applies the unitary transformation described by the displacement
operator $\hat{D}(x,p)=e^{i(p\hat{x}-x\hat{p})}$
to the system $B$.
Bob finally obtains the quantum state $\hat{\rho}^{B}$
which is, in an ideal case, identical to the quantum state
$\hat{\rho}^{Q}$ to be teleported.

The main result of this letter is that the completely positive map
$\hat{\mathcal{L}}$ which describes the continuous variable
teleportation by means of an arbitrary entangled state
and the standard protocol is given by
\begin{equation}
	\hat{\mathcal{L}}\hat{\rho}
	=\int_{-\infty}^{\infty}dx\int_{-\infty}^{\infty}dp\,
	\mathcal{P}(x,p)\hat{D}(x,p)\hat{\rho}\hat{D}^{\dagger}(x,p),
\label{eq-l1}
\end{equation}
with
\begin{equation}
	\mathcal{P}(x,p)=\langle\Psi^{AB}(x,p)\vert\hat{W}^{AB}
	\vert\Psi^{AB}(x,p)\rangle,
\label{eq-l2}
\end{equation}
\begin{equation}
	\vert\Psi^{AB}(x,p)\rangle
	=[\hat{1}^{A}\otimes
	\hat{D}^{B}(x,p)]\vert\Phi^{AB}\rangle,
\end{equation}
where $\vert\Phi^{AB}\rangle$ is a completely entangled state
of continuous variable
\begin{equation}
	\vert\Phi^{AB}\rangle=\frac{1}{\sqrt{2\pi}}\int_{-\infty}^{\infty}
	dx\,\vert x^{A}\rangle\otimes\vert x^{B}\rangle,
\label{eq-l3}
\end{equation}
which satisfies the relations,
\begin{eqnarray}
	\text{Tr}_{A}\vert\Phi^{AB}\rangle\langle\Phi^{AB}\vert
	&=&\frac{1}{2\pi}\hat{1}^{B},\\
	\text{Tr}_{B}\vert\Phi^{AB}\rangle\langle\Phi^{AB}\vert
	&=&\frac{1}{2\pi}\hat{1}^{A}.
\end{eqnarray}
In this state, the systems $A$ and $B$ have the same values of position
and the opposite values of momentum,
\begin{equation}
	(\hat{x}^{A}-\hat{x}^{B})\vert\Phi^{AB}\rangle
	=(\hat{p}^{A}+\hat{p}^{B})\vert\Phi^{AB}\rangle=0.
\end{equation}
When $\mathcal{P}(x,p)=P(x)P(p)$ and $P(x)$ is a Gaussian
function of $x$, the completely positive map $\hat{\mathcal{L}}$
becomes the thermalizing quantum channel.
Hence we refer to the completely positive map $\hat{\mathcal{L}}$
as a generalized thermalizing quantum channel.
In a discrete quantum teleportation of a $N$-dimensional Hilbert space,
this corresponds to a generalized depolarizing quantum channel
obtained by Bowen and Bose \cite{bowen2001b}.
When a pure state $\hat{\rho}=\vert\psi\rangle\langle\psi\vert$
is teleported, the fidelity $F_{\psi}$ is calculated by the formula
\begin{equation}
	F_{\psi}=\int_{-\infty}^{\infty}dx\int_{-\infty}^{\infty}dp\,
	\mathcal{P}(x,p)\vert\langle\psi\vert\hat{D}(x,p)
	\vert\psi\rangle\vert^{2}.
\end{equation}

The property of the simultaneous measurement of position and momentum
is essential for deriving the completely positive map $\hat{\mathcal{L}}$.
The simultaneous measurement which yields the measurement outcome
$(x,p)$ is described by a projection operator,
\begin{equation}
	\hat{X}^{AB}(x,p)=\vert\Phi^{AB}(x,p)\rangle
	\langle\Phi^{AB}(x,p)\vert,
\end{equation}
where $\vert\Phi^{AB}(x,p)\rangle$ is given by
\begin{eqnarray}
	\vert\Phi^{AB}(x,p)\rangle&=&\frac{1}{\sqrt{2\pi}}
	\int_{-\infty}^{\infty}dy\,\vert x^{A}+y^{A}\rangle
	\otimes\vert y^{B}\rangle\,e^{ipy} \nonumber\\
	&=&\left[\hat{D}^{A}(x,p)\otimes\hat{1}^{B}\right]
	\vert\Phi^{AB}\rangle\,e^{-\frac{1}{2}ipx} \nonumber\\
	&=&\left[\hat{1}^{A}\otimes\hat{D}^{B}(-x,p)\right]
	\vert\Phi^{AB}\rangle\,e^{-\frac{1}{2}ipx} \nonumber\\
	&=&\vert\Psi^{AB}(-x,p)\rangle\,e^{-\frac{1}{2}ipx},
\label{eq-b1}
\end{eqnarray}
where $\hat{D}(x,p)$ is the displacement operator,
\begin{equation}
	\hat{D}(x,p)=\exp[i(p\hat{x}-x\hat{p})]
	=\exp(\alpha\hat{a}^{\dagger}-\alpha\hat{a}),
\end{equation}
with $\hat{a}=(\hat{x}+i\hat{p})/\sqrt{2}$ and
$\alpha=(x+ip)/\sqrt{2}$.
The vector $\vert\Phi^{AB}(x,p)\rangle$ is the simultaneous
eigenstate of position-difference operator $\hat{x}^{A}-\hat{x}^{B}$
and momentum-sum operator $\hat{p}^{A}+\hat{p}^{B}$,
which satisfies the relations,
\begin{equation}
	\langle\Phi^{AB}(x,p)\vert\Phi^{AB}(x',p')\rangle
	=\delta(x-x')\delta(p-p'),
\end{equation}
\begin{equation}
	\int_{-\infty}^{\infty}dx\int_{-\infty}^{\infty}dp\,
	\vert\Phi^{AB}(x,p)\rangle\langle\Phi^{AB}(x,p)\vert
	=\hat{1}^{A}\otimes\hat{1}^{B}.
\end{equation}
For quantum optical systems,
the simultaneous measurement of position and momentum
is implemented by a heterodyne detection \cite{leon1997}.

Let us now derive the main result given
by Eqs (\ref{eq-l1})--(\ref{eq-l3}).
Suppose that Alice and Bob share an arbitrary entangled state $\hat{W}^{AB}$
and Alice has a quantum state $\hat{\rho}^{Q}$
which is to be teleported to Bob.
The total quantum state $\hat{\rho}^{QAB}$ of Alice and Bob is given by
\begin{equation}
	\hat{\rho}^{QAB}=\hat{\rho}^{Q}\otimes\hat{W}^{AB}.
\end{equation}
Alice performs the simultaneous measurement of position and momentum
of the compound system $Q+A$, described by the projection operator
$\hat{X}^{QA}(x,p)$, and informs Bob of the measurement
outcome $(x,p)$.
Then Bob applies the unitary operator $\hat{D}^{B}(x,p)$ to the system $B$
and he finally obtains the quantum state,
\begin{widetext}
\begin{equation}
	\hat{\rho}^{B}(x,p)=\frac{\hat{D}^{B}(x,p)
	\left\{\text{Tr}_{QA}\left[
	\left(\hat{X}^{QA}(x,p)\otimes\hat{1}^{B}\right)
	\left(\hat{\rho}^{Q}\otimes\hat{W}^{AB}\right)
	\right]\right\}\hat{D}^{B\,\dagger}(x,p)}
	{\text{Tr}_{QAB}\left[\left(\hat{X}^{QA}(x,p)\otimes\hat{1}^{B}\right)
	\left(\hat{\rho}^{Q}\otimes\hat{W}^{AB}\right)\right]},
\label{eq-l4}
\end{equation}
with probability,
\begin{equation}
	P(x,p)=\text{Tr}_{QAB}
	\left[\left(\hat{X}^{QA}(x,p)\otimes\hat{1}^{B}\right)
	\left(\hat{\rho}^{Q}\otimes\hat{W}^{AB}\right)\right].
\label{eq-l5}
\end{equation}
In deriving Eq.\ (\ref{eq-l4}), we have used the state-reduction formula
\cite{d76,k83,o84}.
Hence the teleported quantum state $\hat{\rho}^{B}_{\mathrm{out}}$ of Bob
is given, in average, by
\begin{equation}
	\hat{\rho}^{B}_{\mathrm{out}}
	=\int_{-\infty}^{\infty}dx\int_{-\infty}^{\infty}dp\,
	\hat{D}^{B}(x,p)\left\{\text{Tr}_{QA}\left[
	\left(\hat{X}^{QA}(x,p)\otimes\hat{1}^{B}\right)
	\left(\hat{\rho}^{Q}\otimes\hat{W}^{AB}\right)
	\right]\right\}\hat{D}^{B\,\dagger}(x,p),
\label{eq-l6}
\end{equation}
which determines the completely positive map $\hat{\mathcal{L}}$
of the continuous variable teleportation by means of
an arbitrary entangled state and the standard protocol.

To obtain the completely positive map $\hat{\mathcal{L}}$
from Eq.\ (\ref{eq-l6}), we expand the entangled quantum state $\hat{W}^{AB}$
in terms of $\vert\Phi^{AB}(x,p)\rangle$'s as
\begin{equation}
	\hat{W}^{AB}=\int_{-\infty}^{\infty}dx
	\int_{-\infty}^{\infty}dy\int_{-\infty}^{\infty}du
	\int_{-\infty}^{\infty}dv\,F_{W}(x,u;y,v)
	\vert\Phi^{AB}(x,u)\rangle\langle\Phi^{AB}(y,v)\vert,
\end{equation}
where we set
$F_{W}(x,u;y,v)=\langle\Phi^{AB}(x,u)\vert\hat{W}^{AB}\vert
\Phi^{AB}(y,v)\rangle$.
Note that the state vector $\vert\Phi^{AB}(x,p)\rangle$ satisfies the relation,
\begin{equation}
	\langle\Phi^{AB}(x,u)\vert\Phi^{BC}(y,v)\rangle
	=\frac{1}{2\pi}\int_{-\infty}^{\infty}dz\,
	\vert z^{C}-y^{C}\rangle\langle z^{A}+x^{A}\vert\,
	e^{-iuz+iv(z-y)}.
\end{equation}
Substituting these equations into Eq.\ (\ref{eq-l6})
and using $\vert x+y\rangle=e^{-iy\hat{p}}\vert x\rangle$,
after some calculation, we obtain
\begin{equation}
	\hat{\rho}_{\mathrm{out}}^{B}=\int_{-\infty}^{\infty}dx
	\int_{-\infty}^{\infty}dp\,F_{W}(x,p;x,p)
	\hat{D}^{B}(-x,p)\hat{\rho}^{B}\hat{D}^{B\,\dagger}(-x,p).
\end{equation}
where $\hat{\rho}^{B}$ is the quantum state of the system $B$,
which is identical to that described by $\hat{\rho}^{Q}$.
Using Eq.\ (\ref{eq-b1}), we can arrive at the result given by
Eqs.\ (\ref{eq-l1})--(\ref{eq-l3}).
Therefore, we have found that the continuous variable teleportation
by means of an arbitrary entangled state and the standard protocol
is equivalent to the generalized
thermalizing quantum channel.

The continuous variable teleportation is usually performed
by means of a two-mode squeezed-vacuum state
$\vert\Psi_{\mathrm{SV}}^{AB}\rangle
=e^{r(\hat{a}^{\dagger}\hat{b}^{\dagger}-\hat{a}\hat{b})}
\vert 0^{A},0^{B}\rangle$ \cite{braunstein1998,furusawa1998}.
In this case, calculating $\mathcal{P}(x,p)
=\vert\langle\Psi^{AB}_{\mathrm{SV}}\vert\Phi^{AB}(x,p)\rangle
\vert^{2}$, we obtain
\begin{equation}
	\hat{\mathcal{L}}\hat{\rho}=\frac{1}{2\pi\bar{n}_{r}}
	\int_{-\infty}^{\infty}dx\int_{-\infty}^{\infty}dp\,
	\exp\left(-\frac{x^{2}+p^{2}}{2\bar{n}_{r}}\right)
	\hat{D}(x,p)\hat{\rho}\hat{D}^{\dagger}(x,p),
\label{eq-t1}
\end{equation}
\end{widetext}
where $\bar{n}_{r}=e^{-2r}$ and $r$ is the squeezing parameter
of the two-mode squeezed-vacuum state
$\vert\Psi_{\mathrm{SV}}^{AB}\rangle$.
The completely positive map $\hat{\mathcal{L}}$ given by Eq.\ (\ref{eq-t1})
is equivalent to the transfer-operator representation
of the continuous variable teleportation \cite{hofmann}.
When we denote the Glauber-Sudarshan $P$-function
of the quantum state $\hat{\rho}$
as $P(\alpha)$, Eq.\ (\ref{eq-t1}) can be rewritten as
\begin{equation}
	\hat{\mathcal{L}}\hat{\rho}=\int\frac{d^{2}\alpha}{\pi}
	P(\alpha)\hat{D}(\alpha)\hat{\rho}_{\mathrm{th}}
	\hat{D}^{\dagger}(\alpha),
\label{eq-t2}
\end{equation}
where $\hat{\rho}_{\mathrm{th}}$ is the thermal state
\begin{equation}
	\hat{\rho}_{\mathrm{th}}=\frac{1}{1+\bar{n}_{r}}
	\sum_{n=0}^{\infty}\left(\frac{\bar{n}_{r}}{1+\bar{n}_{r}}\right)^{n}
	\vert n\rangle\langle n\vert.
\end{equation}
This result explicitly shows that the continuous variable teleportation
by means of the two-mode squeezed-vacuum state is nothing but
the thermalizing quantum channel.
When the two-mode squeezed-vacuum state is shared through
a noisy quantum channel,
the parameter $\bar{n}_{r}=e^{-2r}$ appeared
in Eq.\ (\ref{eq-t1}) is replaced with
$\bar{n}_{r}=1-\left(1-e^{-2r}\right)T$
\cite{Lee2000,chizhov,takeoka},
where $T$ stands for the transmission coefficient
of the noisy quantum channel.

We next consider the continuous variable quantum dense coding
by means of an arbitrary entangled state and the standard protocol
in which Alice applies the unitary operator $\hat{D}(x,p)$
to encode some classical information
and Bob performs the simultaneous measurement of position
and momentum to extract the information
\cite{ban1999,braunstein1999,ban2000a,ban2000b}.
Suppose that Alice and Bob share an arbitrary entangled quantum
state $\hat{W}^{AB}$ to transmit classical information
via the continuous variable quantum dense coding.
When Alice applies the unitary operator $\hat{D}(x,p)$
to her system and sends it to Bob,
his quantum state $\hat{W}^{AB}(x,p)$ is given by
\begin{widetext}
\begin{equation}
	\hat{W}^{AB}(x,p)=\int_{-\infty}^{\infty}d y
	\int_{-\infty}^{\infty}dz\int_{-\infty}^{\infty}du
	\int_{-\infty}^{\infty}dv\,F_{W}(y,u;z,v)
	\vert\Phi^{AB}(x+y,p+u)\rangle
	\langle\Phi^{AB}(x+z,p+v)\vert.
\end{equation}
Thus the conditional probability (the channel matrix of the quantum
dense coding) $P(x',p'\vert x,p)$ that Bob obtains the measurement
outcome $(x',p')$ when Alice encoded $(x,p)$ is calculated to be
\begin{eqnarray}
	P(x',p'\vert x,p)&=&\langle\Phi^{AB}(x'-x,p'-p)\vert
	\hat{W}^{AB}\vert\Phi^{AB}(x'-x,p'-p)\rangle \nonumber\\
	&=&F_{W}(x'-x,p'-p;x'-x,p'-p) \nonumber\\
	&=&\mathcal{P}(x-x',p'-p),
\end{eqnarray}
where $\mathcal{P}(x,p)$ is given by Eq.\ (\ref{eq-l2}).
When $\hat{W}^{AB}$ is the two-mode squeezed-vacuum state,
we obtain
\begin{equation}
	P(x',p'\vert x,p)=\frac{1}{2\pi\bar{n}_{r}}\exp\left[
	-\frac{(x'-x)^{2}+(p'-p)^{2}}{2\bar{n}_{r}}\right].
\label{eq-d1}
\end{equation}
\end{widetext}
In the strong squeezing limit $(r\gg1)$,
the channel matrix $P(x',p'\vert x,p)$ is approximated
as $\delta(x'-x)\delta(p'-p)$ and thus the continuous variable
quantum dense coding can transmit the amount of classical
information twice as that without quantum entanglement
\cite{braunstein1999}.
When the two-mode squeezed-vacuum state is sent through a noisy
quantum channel,
the parameter $\bar{n}_{r}=e^{-2r}$ appeared
in Eq.\ (\ref{eq-d1}) is replaced with
$\bar{n}_{r}=1-\left(1-e^{-2r}\right)T$
\cite{ban2000a,ban2000b}.

In summary, we have shown that the continuous variable quantum
teleportation by means of an arbitrary entangled quantum state
and the standard protocol is equivalent to the generalized
thermalizing quantum channel.
This is a continuous version of the result obtained
by Bowen and Bose \cite{bowen2001b}, in which they have shown
that the discrete (the finite dimensional) quantum teleportation
with the standard protocol is equivalent to
the generalized depolarizing quantum channel.
In particular, when a two-mode squeezed-vacuum state
is used as an entanglement resource,
the continuous variable quantum teleportation becomes
the thermalizing channel in which the number of thermal
photons is given by $\bar{n}_{r}=e^{-2r}$.
The continuous variable quantum dense coding by means of
an arbitrary entangled state the standard
protocol has been also considered.
As long as the standard protocol is applied,
not only the continuous variable quantum teleportation
but also the continuous variable quantum dense coding
is completely determined by the function $\mathcal{P}(x,p)$
given by Eq.\ (\ref{eq-l2}).

\bibliographystyle{apsrev}
\bibliography{th-channel}
\end{document}